# Sequestration of atmospheric carbon dioxide as inorganic carbon in the unsaturated zone under semi-arid forests

Israel Carmi[1], Joel Kronfeld[1], Murray Moinester[2]

[1]Department of Geosciences, Tel Aviv University, 69978 Tel Aviv, Israel, carmiisr@post.tau.ac.il, joel.kronfeld@gmail.com

[2]School of Physics and Astronomy, Tel Aviv University, 69978 Tel Aviv, Israel, murray.moinester@gmail.com

## Abstract

Inorganic carbon, in the form of allogenic (transported) and pedogenic (soil) carbonates in semi-arid soils, may comprise an important carbon sink. Carbon dioxide, $CO_2$, originating from the atmosphere and exhaled by tree roots into the soil, may be hydrated by soil water within the unsaturated zone (USZ) of semi-arid soils to produce the carbonic acid ($H_2CO_3$) solutes $HCO_3^-$ bicarbonate and $H^+$ Hydrogen ion. This $H^+$ may then dissolve relict soil $CaCO_3$ carbonate (calcite), to release $Ca^{+2}$ calcium cations and more $HCO_3^-$ bicarbonate. When conditions allow, one mole of $Ca^{+2}$ and two moles of $HCO_3^-$ combine to precipitate one mole of calcite, and to release one mole of $CO_2$: $Ca^{+2} + 2HCO_3^- \rightarrow CaCO_3\downarrow + CO_2\uparrow + H_2O$. However, it has been claimed that such carbonates do not sequester significant amounts of present day atmospheric $CO_2$. The reasons given were that they originate in part from the pre-existing limestone; and that for every mole of calcite precipitated, one mole of $CO_2$ may be liberated to the atmosphere. It was argued that only if the $Ca^{+2}$ cation is derived from a non-carbonate source can sequestration be assumed. We have tested these assumptions under field conditions at two semi-arid sites in Israel. We found that bicarbonate, originating from root exhalation, is depleted and is incorporated within the USZ as carbonates precipitate. Thus, a net sequestration of atmospheric $CO_2$ does occur under semi-arid forests. Moreover, most of the $CO_2$ liberated in the precipitation reaction may remain in the soil. And $Ca^{+2}$ in the sediment may also be supplied from sources other than pre-existing calcite. Forestation can therefore augment pedogenic carbonate formation. By extrapolating our data globally, we suggest that worldwide semi-arid forests (existing and to be planted) may sequester 5-20% of the current annual anthropogenic increase of atmospheric carbon dioxide as pedogenic carbonate.

## 1. Introduction

Since the Industrial Revolution, the carbon dioxide concentration in the atmosphere has risen from approximately 280 ppmv (parts per million by volume) to approximately 405 ppmv at present (Craig and Keeling, 1963; ProOxygen, 2017). The present global atmospheric $CO_2$ reservoir of ~3150 billion tons has been recently increasing annually by ~20 billion tons. This increase has occurred predominantly through the burning of fossil fuels, and secondarily through the processes of deforestation and desertification and forest fires. This is releasing carbon that had been previously taken out of the atmosphere in prior geological eras, particularly during the Carboniferous Period, when the great coal deposits were laid down (Berner and Kothavala, 2001; Rothman, 2001, 2002; Bergman et al., 2004; Franks et al., 2014). The $CO_2$ had been stored primarily as a chemically reduced form of plant-based organic material that had been converted to coal or petroleum. These fuels are now being oxidized and the carbon is being returned to the atmosphere as carbon dioxide. Considering the goal of limiting global warming, large climate engineering projects have been proposed; for example, carbon dioxide removal (CDR) and solar radiation management (SRM) (Linnér and Wibeck *(2015)*). The effectiveness, cost and risk of these and other proposals are currently being investigated. Some of these methods should indeed be employed to retard the increasing atmospheric $CO_2$ concentration. Another way being considered is to once again utilize plant photosynthesis to abstract atmospheric $CO_2$, and store it as organic carbon in trees. Thus, planting trees has been proposed as being effective in sequestering atmospheric $CO_2$, both as Above-ground Biomass Carbon (ABC) as well as in the roots (Johnson and Colburn, 2010; Kell, 2012; Tans and Wallace, 1999; Watson et al., 2000). The carbon observed in the above ground plant mass, the leaf litter, and organics disseminated in the soils (soil organic matter), comprise organic carbon. Forestation can store large amounts of organic carbon. For example, almost 4 billion tons of ABC has been stored since 2003 by tree planting in northern China. This equals almost 25% of its annual fossil fuel emissions (Liu et al., 2015). Ancillary benefits also accrue. Among these are soil stabilization, reduced erosion and runoff, improved soil structure and quality, as well as reduced soil biogenic nitric oxide (NO) emissions (Gelfand et al., 2009). However, all may not be so sanguine, considering that trees have a finite lifetime. Except for exceptional species, they should be considered as carbon stores in terms of centuries and not millennia. When trees die and decompose, much of the carbon recycles as $CO_2$. Moreover, most forestation efforts have so far been carried out in temperate zones (Martin et al, 2001). However, this is where most productive agriculture is carried out. Boysen points out (2015) that attaining mandated climate goals via ABC would require cutting food production by converting croplands to forests. Besides, large amounts of fertilizer would be required whose runoff would likely degrade water supplies. Thus, ABC conversion of atmospheric $CO_2$ to terrestrial biomass may come at an egregious cost. Planting forests in the temperate zone is not the optimum model. Investigated here is the proposal to sequester atmospheric carbon dioxide through forestation in semi-arid regions particularly as inorganic pedogenic carbonate beneath the ground surface within the unsaturated zone (USZ).

Semi-arid regions have marginal agriculture, but indigenous flora (trees and shrubs with deep roots) can be sustained by the approximately 250-600 mm of annual rainfall. In these regions, the unsaturated zone (USZ) is considerably more extensive than in temperate regions. It extends from the land surface cover downwards through the sediment to the water table; and has the potential for long-term carbon sequestration.

The relative effective thicknesses of the USZs in different climatic regions can be readily seen by comparing their average maximum rooting depths. For temperate regions, the maximum rooting depths of coniferous and deciduous forests are ~2.8 m. In drier areas, plants sink roots much deeper to reach water, in order to survive. The roots of the Shepherd (*Boscia albitrunca*) and Acacia trees can exceed 60 m in Africa's Kalahari, while the Israeli Tamarisk (*Tamarix aphylla*) can send its roots down 20 m (Canadell et al., 1996). The process of long-term storage of inorganic carbon below the ground surface has generally been neglected. At best, it was studied to only a maximum depth of one meter, and without the benefit of isotopic tracers. Yet, this process can be more important than the above ground organic carbon storage, when carbon sequestration over the long term is considered. The $CO_2$ that is released through the roots, combined with bacterial oxidation of soil organic matter, can generate soil $CO_2$ partial pressures that can be orders of magnitude higher than its value (405 ppmv, 4.05 x$10^{-4}$ atm) in the overlying ambient atmosphere (Carmi et al., 2013; Clark and Fritz, 1999; Pumpanen et al., 2008). Some of this gas diffuses upwards and leaves the soil (Carmi et al., 2015). The rest can interact with soil moisture. The high partial pressure of $CO_2$ facilitates the formation of dissolved inorganic carbon (DIC), which comprises the sum of all inorganic carbon species in a solution: $CO_2$, carbonic acid, bicarbonate anion, and carbonate. The DIC later combines with appropriate bivalent cations and precipitates, in turn, as pedogenic carbonate at varying depths within the USZ. Throughout the discussion, calcium represents all of the other cations, (e.g. iron and magnesium), that can form carbonates. Where rainfall is plentiful, this precipitate dissolves. In semi-arid regions, where rainfall is sparse, precipitated calcite (pedogenic carbonate) can remain stable for thousands of years (Cerling, 1984). The $CO_2$ that remains dissolved as bicarbonate percolates downwards and eventually enters the aquifer. Thus, once precipitated within the USZ or upon entering the aquifer, appreciable amounts of atmospheric $CO_2$ are stored as $CaCO_3$ within the USZ or as $HCO_3^-$ in the ground water of the aquifer beneath the USZ, for millennia. The pedogenic carbonate formed in the USZ and the bicarbonate (DIC) together comprises sequestered inorganic carbon. Note however that besides cooling via $CO_2$ removal, forests warm due to reduced albedo and cool due to evapotranspiration. All such effects should be included when evaluating the possible net climate benefits of forestation due to carbon sequestration (Rotenberg & Yakir, 2010, 2011; Tal & Gordon, 2010).

## 2. Theoretical considerations of the role inorganic carbon that plays in the sequestration of atmospheric CO$_2$ in the USZ.

Soils are a sink for carbon: both organic and inorganic carbon can be sequestered. Inorganic carbon stocks can be quite considerable; yet, their importance in sequestering atmospheric derived CO$_2$ has been largely neglected (Schlesinger, 2000; Chang, 2002). Monger (2014) presents several cogent reasons for this contention, which are addressed below. Foremost, he believes that for pedogenic carbonate to be a net sink, the calcium ions must come from silicate weathering, or some other non-carbonate source. Moreover, he points out that it is technically difficult to determine readily the ratio of lithogenic to pedogenic carbonate in a soil. Inorganic differentiation of carbon takes on two basic types, with different relevancy as to the ability to abstract and store atmospheric carbon. Carbonates are common, particularly in semi-arid soils. Here they are more stable than in wetter temperate regions, where they tend to dissolve. Lithogenic carbonates, primarily marine limestone, were formed in the distant past. Their carbonate content bears no relationship to the present atmosphere. Pedogenic carbonate is authigenic. It forms within the soil profile as DIC combines with calcium cations in the soil moisture, eventually precipitating as calcite. Trees and other rooting plants take in atmospheric carbon dioxide through photosynthesis, eventually exhaling carbon dioxide through their roots into the unsaturated zone of the soil. The partial pressure of the soil gas is much greater than that of the ambient atmosphere. The carbon isotopic ratios are related to the photosynthetic pathway of the plant. C3-type plants (whose $\delta^{13}$C averages -26‰) exclusively dominate in the study areas; while C4 plants, which yield much more enriched $\delta^{13}$C values, are not represented there (Vogel et al., 1986). The CO$_2$ exhalations are thus strongly isotopically depleted compared to the CO$_2$ in the present atmosphere ($\delta^{13}$C of -9 ‰). The $\delta^{13}$C(‰) is the relative deviation of the $^{13}$C/$^{12}$C ratio of the sample compared to that of a standard material, reported in parts per thousand (Stuiver and Polach,1977). The standard is a marine fossil from the Pee Dee Formation in South Carolina. A series of fractionation steps (e.g., Clark and Fritz, 1999, p.120) affect the carbon isotopic ratios: CO$_2$ gas root exhalation, CO$_2$ (g) dissolving and converting to aqueous CO$_2$ (aq), formation of HCO3$^-$ bicarbonate anions, precipitation of the solid carbonate. In the transformation of CO$_2$ (g) to the solid carbonate precipitate, while the isotopic difference between the sediment and the atmosphere is slight for $^{13}C$ (~1%), it is 100% for $^{14}C$. The host sediment contains no $^{14}C$, which is created only in the upper atmosphere.

The relevant chemical equations for the process of the dissolution of the soil gas into the soil moisture, the formation of carbonic acid and the precipitation of calcite can be presented in simplified form as:

1) $CO_2 + H_2O \rightarrow H_2CO_3 \rightarrow H^+ + HCO_3^-$

Soil-gas $CO_2$ combines with the soil moisture to form a carbonic acid solution, which rapidly dissociates to $H^+$ and $HCO_3^-$. In this reaction, atmospherically derived $CO_2$ becomes bicarbonate (DIC), generally the most common anion of ground water. If the DIC enters the water table, it can be stored in the groundwater long-term, dependent on the age and flow rate. In Israel, like other semi-arid zones, the flow rate is slow, and the water can be thousands of years old (Kronfeld et al., 1993; Rogojin et al., 2002). The dissolution reaction (Eq. 1) is comparable to carbon sequestration, as the source is atmospheric $CO_2$ (e.g., Suarez, 2000; Drees et al., 2001; Monger et al., 2015).

2) $CaCO_3 + H^+ \rightarrow Ca^{2+} + HCO_3^-$

The carbonic acid, formed in Eq. 1, can dissolve existing calcite (dominantly limestone) within the soil to release calcium ions. The $CO_3$ in Eq. 2 is from relict carbonate, not related to modern atmospheric $CO_2$. This $HCO_3^-$ does not therefore sequester atmospheric $CO_2$.

3) $Ca^{2+} + 2HCO_3^- \rightarrow CaCO_{3\downarrow} + CO_{2\uparrow} + H_2O$

The cations released in Eq. 2 combine with the DIC. Facilitated by various processes, principally degassing of $CO_2$ or the evaporation of $H_2O$, secondary calcite precipitates. However, Eq. 3 would imply that no net $CO_2$ is sequestered, if for every mole of calcite formed, one mole of $CO_2$ returns to the atmosphere. Thus, it has been assumed (e.g., Monger et al., 2015) that the formation of pedogenic carbonate is neutral as to atmospheric carbon abstraction and storage. However, when this reaction occurs within the USZ soil column, generally thick in semi-arid regions, only the topmost fraction is in direct contact with the atmosphere. Though this may be considered an "open system", the rate of release of the gas is limited by the rate of diffusion. Much of the $CO_2$ released by Eq. 3, particularly from precipitation occurring at depth, should be expected to enter and mix with the relatively high partial pressure $CO_2$ in the USZ soil-gas

If there are non-carbonate sources of supply of calcium ions, net atmospheric $CO_2$ sequestration will occur. For instance, this can be seen in the Eq. 4 weathering of a common Ca bearing silicate, wollastonite.

4) $2CO_2 + 3H_2O + CaSiO_3 \rightarrow H_4SiO_4 + 2HCO_3^- + Ca^{+2}$

In the weathering of silicates, all of the $CO_2$, unlike in carbonate weathering, is derived ultimately from atmospheric $CO_2$. In this representative reaction, two moles of atmospherically derived $CO_2$ are used to produce two moles of $HCO_3^-$ but only one mole of $CO_2$ is liberated when these two moles combine with $Ca^{+2}$ to precipitate $CaCO_3$, as in Eq. 3. A net sequestration of atmospheric $CO_2$ therefore unambiguously occurs (Kolosz et al., 2016; Kolosz et al., 2017; Hartmann et al., 2017; Washbourne et al., 2015). A variety of sources of calcium ions, outside silicate weathering or from relict limestone, are present in nature. Moreover, if there are other sources of calcium in the sediment, which can be adsorbed as a cation on the surface of admixed clay minerals, weathered feldspar ($CaAl_2Si_2O_8$), or gypsum ($CaSO_4 \cdot 2H_2O$), sea spray or rain, then there may be net sequestration of atmospheric $CO_2$ when $Ca^{+2}$ combines with $HCO_3^-$ (Eq.1) to precipitate calcite. It is important therefore to evaluate for sequestration under the field conditions rather than assuming the theoretical or laboratory conditions. Singer (2007) has shown that the dominant adsorbed cations on the clays of Israel are calcium and magnesium. Moreover, these cations are readily amenable to being replaced by $H^+$ ions from carbonic acid (Eq.1). For areas relatively close to the sea, the cations may be introduced by sea spray and rain. Loewengart (1961) has shown that aerosols derived from sea spray are the major source of salinity for rainwater recharge in Israel.

Two previous field studies in Israel indicated that precipitation within the USZ is occurring, and that there is a net transfer of atmospheric $CO_2$ when calcite is precipitated (Carmi et al., 2009; 2013, 2015). The primary purpose of these investigations was to study the evolution of carbon isotopes as carbon (in the gas, liquid and solid phases) traverses the USZ to the water table. These studies are now revisited, with the addition of some new data; for they provide a new perspective, as to whether or not there is a net storage of atmospheric $CO_2$ within the soil profile as pedogenic calcite. This aspect has not been previously considered. To assist in tracing the source of the carbon dioxide, [13]C and [14]C isotopic abundances were measured as well. The latter is a unique tracer for labeling atmospheric derived sources of gas, liquid and solid phase carbon within the USZ.

## 3. Nizzanim (Nitzanim) and Yatir sampling sites reconsidered in view of the inorganic sequestration of atmospheric $CO_2$

### 3.1 Sampling

Two sampling sites, both within the semi-arid USZ, were studied. The first was located in the USZ above the Coastal Plain aquifer within the non-irrigated part of the Nizzanim Nature Reserve (GPS: 31.773016, 34.655555). The sands of the USZ are of Pleistocene–Holocene age. They are composed of quartz, carbonate, and clay with occasional additions from dust storms. The other sampling site was located in the Yatir Forest (GPS: 31.347129, 35.050898), situated above the Mountain Aquifer (also called the Judea Group Aquifer, a carbonate aquifer) at an elevation of ~650 m above sea level, along the southwestern flanks of the Judean Hills, at the edge of Israel's Negev desert. Beginning in 1966, Keren Kayemeth LeIsrael - Jewish National Fund foresters planted some million trees at Yatir (KKL-JNF, 2016). It is now the largest forest in Israel, covering 28 km². The topsoil is rather shallow and is composed of loess, mostly quartz, with carbonate minerals and clays, as well as a lesser percentage of feldspar minerals. This aeolian deposit, transported from the Sahara and the Arabian Deserts, blankets the underlying carbonate rocks. The forest comprises mainly Aleppo pine (*Pinus halepensis* Mill), with some subordinate cypress and other pine species. The trees have a mean height of 11 m; the tree density is approximately 30,000 $km^{-2}$, and the average depth of the soil cover is generally less than ~1 m. The understory vegetation is minor. The annual precipitation of approximately ~30 cm, mostly falling during the winter as high intensity rain events, makes Yatir one of the world's driest forests. Below the USZ of Yatir is an impermeable carbonate stratum (aquiclude), about 300 m thick. Despite the present low precipitation, this new forest is productive and stores carbon relatively effectively (Grünzweig et al., 2003).

Semi-arid regions, having a significant global surface area, have the potential to store carbon in forested biomass and soil (Grünzweig et al., 2003; Grünzweig et al., 2007; Rotenberg and Yakir, 2010, 2011). Savannas have been shown to be carbon sinks (Fisher et al., 2014), especially when the grasslands are converted to forests (Hibbard et al., 2001, 2003). The effectiveness of the carbon sequestration increases significantly when the annual precipitation is below 350 mm/yr (Jackson et al., 2002). In all of the above cases, only *organic carbon* was considered. The importance of *inorganic carbon* for sequestering atmospheric $CO_2$, coupled to the unsaturated zone of semi-arid regions, has been only recently been stressed (Moinester et al., 2014; Godfrey-Smith et al., 2015; Moinester et al., 2016).

## 3.2 Sample Collection, Preparation and Isotopic Analyses

The Nizzanim section was sampled from the ground surface down to the water table, which is currently 22 m below the surface. In May 2008, sediment samples were collected from two drilled soil profiles in the Yatir forest. Data for a 4.5 meter thick section were taken in a bare plot ~5 m from the forest canopy, an area containing tree roots. A shallower profile reaching to 90 cm was also sampled in a plot directly under the forest canopy, an area containing roots from seasonal plants, but not from tree roots. Results of this latter profile are presented here for the first time (Table 1). In the field, following collection, the sediments were packed in two tightly closed plastic bags, one inside the other, and then refrigerated at 5° C. Processing of the samples was carried out at the Weizmann Institute's Kimmel Center for Archaeological Science. The moisture content of the sediment, and the $CO_2$ derived from the dissolved inorganic carbon (DIC) species in the soil water, were obtained by vacuum distillation to total dryness (Carmi et al., 2007, 2009).  The DIC concentration of the soil water (mmol $CO_2$ $L_w^{-1}$) was calculated by dividing the amount of collected $CO_2$ by the amount of collected water ($L_w$). The DIC concentration per liter of sediment (L) was determined from measurements of soil porosity and humidity. For example, a liter of Yatir sediment was found to contain 47% solid + 12 % water + 41% gas. The $\delta^{13}C$ (‰) isotopic fraction of the DIC was measured in a Finnigan Mat 250 mass spectrometer, at the Weizmann Institute Department of Environmental Science and Energy Research. The $^{14}C$ in the DIC was converted to graphite, and the radiocarbon activity was measured by Accelerator Mass Spectroscopy at the University of Arizona. The $^{14}$C activity in the inorganic fraction of the sediment was measured in a Wallac 1220 liquid scintillation detector at the Weizmann Institute of Science. In order convert the rates of change in DIC and the carbon isotopes as a function of depth to a function of time, the rates at which soil moisture infiltrated downward were determined using the small component (HTO) of radioactive tritium (12.3 year half-life) in natural rain water as a tracer. Tritium concentrations in water, extracted from the soil profile at Yatir, were measured as a function of depth, also using the Wallac 1220 detector. The rate of water descent was thus determined to be 11 cm/year (Carmi et al., 2015). At Nizzanim, the tritium concentration in the soil profile had a sharp maximum at 18 meters depth. This was interpreted as a remnant of the peak of atmospheric tritium due to atmospheric thermonuclear contamination during 1962-1964 nuclear tests, some 40 years before the collection of the sediments. The flow rate at Nizzanim was thus determined to be 45 cm/yr (1800/40) (Carmi et al. 2009).

## 4. Results and Discussion

### 4.1 The forestation additions of atmospheric derived $CO_2$ stored in the USZ

As soil moisture, charged with DIC generated from the soil gas, percolates slowly downwards at the Nizzanim site, it was noted that calcite precipitation occurs along the pathway of descending recharge within the USZ. The precipitation rate was measured to be 0.071% per cm per liter of sediment (L) with respect to the 1 meter deep level DIC density of ~6.9 mmol $CO_2$ $L^{-1}$. The value 0.071% $L^{-1}$ $cm^{-1}$ corresponds to a rate of decrease in the DIC of ~3.2% $CO_2$ $yr^{-1}$ $L^{-1}$ (0.071% x 45 cm/yr). Fig. 1 (data from Carmi et al. (2009, 2014)) shows that both the DIC concentration and the $^{14}$C activity in the solid pedogenic carbonate decrease as a function of depth.  The $^{14}$C activity is of particular interest, as the host calcite is too old to contribute radiocarbon. The only significant radiocarbon source for the sediment of the USZ is atmospheric $CO_2$, that was injected into the DIC by the overlying plant roots. Once the $^{14}CO_2$ enters the DIC, its activity decreases not as a function of age, for the passage of time is too short compared to the [14]C half-life of 5,730 years. Rather, the decrease in DIC activity is associated with removal of carbon (all three carbon isotopes) by precipitation onto the solid pedogenic carbonate horizons in the USZ, paralleling the path of descending water. The qualitative decrease of [14]C activity in the solid phase (Fig. 1) reflects the fact that with increasing depth, there is continuously less radiocarbon to precipitate from the DIC. What is occurring is the exchange of radiocarbon as the pedogenic carbon is precipitated and partly dissolved, only to be re-precipitated continuously down the section. The incorporation of atmospherically derived radiocarbon into the crystal structure shows that Eqs. 1-3 are not applicable for the "dirty" conditions found in the field. In Israel, there are several potential sources of calcium ions that can combine with the soil gas $CO_2$, ions that were not derived from dissolving of pre-existing carbonate minerals. The sources most likely to supply calcium in our specific case are from desorption from the surface of clay minerals (brought in by airborne dust) on which they were absorbed, and secondarily by rain/sea spray depending on the geographical location with respect to the coast line.

The rate of decrease in DIC is not as readily apparent in the 4.5m section at Yatir as it is at Nizzanim. It appears that in this much condensed section, a tangle of roots are actively respiring $CO_2$ throughout most of the top 2.5 meters of the section. Thus, the DIC does not evince the clear pattern of decreasing DIC down the section. It is rather constant, at least in the upper 2.5 m, averaging 4.5 mmol L$^{-1}$. DIC that is precipitated is rapidly renewed by the high ambient $CO_2$ pressure throughout most of the USZ. The isotopic data presented in Fig. 2 provides firm evidence that exchange with the solid phase is continuous. The $\delta^{13}C$ of the DIC (originating from isotopic depleted $CO_2$ from C3 plants ($\delta^{13}C = \sim$ -26 ‰) is slowly enriched to approximately half this value by exchange with a more carbon-isotopically enriched source within the USZ. The source is the relict marine carbonate having $\delta^{13}C \sim 0$ ‰. The $\delta^{13}C$ of this carbon exchange clearly displays a progressive enrichment in the DIC downward profile.

There is a correlated depletion of the solid from $\delta^{13}C \sim 0$ ‰ to $\delta^{13}C \sim -7$‰. The exchange is actually a continuous process of dissolution of the relict carbonate and precipitation within the USZ.

The $^{14}C$, which has only a single atmospherically derived source, traces the incorporation of the soil-gas derived $CO_2$ dissolved in the DIC into the pedogenic precipitate with depth. The $\Delta^{14}C$ values shows analogously to Nizzanim, though the DIC concentration at Yatir remains constant over most of the interval, that exchange is nonetheless occurring continuously. $\Delta^{14}C$ is the relative deviation of the $^{14}C$ activity in the sample from that of a standard reference, normalized for $^{13}C$ fractionation, but without age correction (Stuiver and Polach,1977). The primary (marine) solid calcite in Yatir has no $^{14}C$, corresponding to $\Delta^{14}C = -1000$ ‰ in Fig. 2. The carbon exchange is clearly noted for example in the two topmost data points, where $\Delta^{14}C$ (‰) $\sim -700$ ‰ in the solid phase is enriched with respect to $\Delta^{14}C = -1000$ ‰. At the same time $\Delta^{14}C$ (‰) in DIC is depleted progressively: $\Delta^{14}C$ (‰) $\sim -100$ ‰ near surface to $\Delta^{14}C$ (‰) $\sim -500$ ‰ lower in the section. Pedogenic carbonate must be precipitating along the path. These data demonstrate that atmospherically derived $CO_2$ is being incorporated into the sediment. The released $CO_2$ (Eq.3) probably does not return to the atmosphere, as the equation would suggest. Rather, the deeper portions of the sections are not open systems, in contrast to the topmost sections that are in contact with the atmosphere. At depth, the released $CO_2$ moves slowly by diffusion, and then enters and mingles with high partial pressure of the soil gas within the USZ. The rate at which the radiocarbon precipitates onto the pedogenic carbonate cannot be precisely quantified, because the process is not a closed system. It merely denotes that atmospheric $CO_2$ is indeed incorporated with pedogenic carbonates and that the carbonates should be a net sink, and not net neutral with regard to $CO_2$ sequestration. It had been previously suggested (Carmi, et al., 2015) that the radiocarbon activity monitored in the sediment was possibly due to having been brought in adsorbed to windblown dust. We know now that is not the case but rather the process of incorporation of radiocarbon is as described here above.

## 4.2 The baseline value of atmospheric-derived $CO_2$ stored in the USZ

In the USZ of Nizzanim and Yatir, the $^{14}C$ exchange rates (Figs. 1,2) from the descending DIC were given as 1.2% $yr^{-1}$ $L^{-1}$ and 1.0 % $yr^{-1}$ $L^{-1}$, respectively (Carmi et al. 2009; 2014). Because these decreases in $^{14}C$ exchange rates were found to be very similar, it was assumed that the carbonate precipitation rates were also similar. The Yatir carbonate precipitation rate was therefore previously assumed to be approximately 2.7% $yr^{-1}$ $L^{-1}$ (3.2%/1.2), equivalent, per liter of sediment, to 5.4 mg $CO_2$ $yr^{-1}$ $L^{-1}$ removed from the atmosphere and stored in the USZ. Precipitation rates however depend on many factors: pH, alkalinity, saturation index, supersaturation, temperature, humidity, the $Ca^{+2}$ concentration, the DIC concentration, pressure, agitation, soil aggregates, etc. Since these factors vary from one forest to another, the preferred value is the one based on precipitation data taken in or near the same forest. To better estimate the carbonate precipitation rate, we now consider a part of the Yatir forest area where the effects of roots from seasonal plants, not trees, appear to be dominant (Table 1). Data from the top 30 cm are not taken into account here for determining the sequestration rate, since the DIC density and moisture so close to the surface are too strongly affected by evaporation, temperature changes, and other variables.

The concentration of the DIC decreases from 4.5 to 3.4 mmol C $L^{-1}$. The The 1.1 mmol C $L^{-1}$ difference in DIC concentration represents precipitation of calcite over 30 cm within the USZ, corresponding to a 0.81% decrease per cm per liter of sediment (0.81 $cm^{-1}$ $L^{-1}$). This value is taken to be characteristic of the forest soil, appropriate to regions of seasonal plants and/or trees. The rate of water descent is 11 cm/yr (Carmi et al., 2014). Thus, we can estimate the yearly inorganic carbon precipitation rate (storage in the USZ) as 8.9% $yr^{-1}$ $L^{-1}$ (0.81x11). The annual deposition rate of inorganic carbonate onto the sediment for the seasonal plants in this section is then 0.40 mmol $CO_2$ $yr^{-1}$ $L^{-1}$ (0.089x4.5), equivalent to 17.6 mg $CO_2$ from the atmosphere per year per liter of sediment. Note however that this rate decreases with depth, by ~81% per meter, since the input source of $CO_2$ is only from roots near the surface.

The soil humidity is between 11 and 12%. The $\delta^{18}O$ in the local rainfall is approximately -5‰ with respect to Vienna Standard Mean Ocean Water (Gat and Dansgaard, 1972). The $\delta^{18}O$ in soil moisture is highly enriched from 0‰ to -2 ‰ due to evaporation, since water vapor formed during evaporation of liquid water is enriched in $H_2^{16}O$, and the residual liquid is therefore enriched in $H_2^{18}O$. The $\delta^{13}C$ in the same samples is typical for bicarbonate which is converted from $CO_2$ exhalation ($^{13}C$-depleted) into the soil from C3 type plants, *in situ.* Thus the recharge water is introduced at and then descends from the ground surface where it had undergone evaporation, while the DIC is generated within the USZ from $CO_2$ exhalations below ground, and does not evince evaporation isotopic effects.

**4.3 Using the initial USZ storage value to estimate the potential for $CO_2$ abstraction for global semi-arid regions**

Though the database is very small, its importance can be seen when extrapolated to the large global areas of semi-arid regions, which contain much thicker USZ sections. In the Loess Plateau of China, for example, the soil layer is ~ 50m thick, though the inorganic and organic carbon has been studied only in the top 1 meter (Chang et al., 2012). For the forest wall of the Sahel (8000 km long by 15 km wide, $1.2 \times 10^5$ km$^2$, (Guardian, 2011; GGW, 2016), a reasonable USZ sequestration estimate can be made, because the mineralogy of the sediment (loess) is similar to that of Yatir. Estimating the average depth of $CO_2$ producing roots to be 6 m, the sequestration volume is $7.2 \times 10^{14}$ liters ($1.2 \times 10^{11} \times 6 \times 10^3$). The inorganic sequestration rate can then be estimated as 12.6 million ton of $CO_2$ per year from the atmosphere ($7.2 \times 10^{14}$ L $\times 17.6 \times 10^{-3}$ g L$^{-1}$ = $12.6 \times 10^{12}$ g). Similar efforts are ongoing in India and China.

It is useful to compare this rate to a baseline rate in which the Sahel area has only seasonal plants rather than trees. The input source of $CO_2$ is then only from roots near the surface. The precipitation rate decreases with depth by 8.9% yr$^{-1}$ from 17.6 mg $CO_2$ yr$^{-1}$ L$^{-1}$ near the surface, corresponding to a decrease of 81% m$^{-1}$, assuming an 11 cm yr$^{-1}$ flow rate. The average precipitation rate over the 6 meter depth is then ~2.2 mg yr$^{-1}$ L$^{-1}$. The inorganic sequestration rate is then estimated as 1.6 million ton of $CO_2$ per year from the atmosphere ($7.2 \times 10^{14}$ L $\times 2.2 \times 10^{-3}$ g L$^{-1}$ = $1.6 \times 10^{12}$ g). That is, the increase in the sequestration rate (from 1.6 to 12.6 M tons yr$^{-1}$) due to forestation is then roughly 11 million ton of $CO_2$ per year.

Extrapolating to the global semi-arid regions, only a *"gedanken"* guestimate is possible. Assume that the topsoil composition and depth and calcite precipitation rate in Yatir is representative of the global semi-arid area [~24 million km$^2$ or 2.4 billion ha], ~17.7% of the global surface area (Lal, 2004) following forestation. Taking 6 m as the average depth of root respiration, the area 24 million km$^2$ gives a sediment volume of $1.44 \times 10^{17}$ liters. Therefore, roughly 2.5 billion tons of $CO_2$ ($17.6 \times 10^{-3}$ g L$^{-1}$ x $1.44 \times 10^{17}$ L ~2.5 Pg) could potentially be precipitated globally each year in the USZ as calcite, following forestation. Uncertainties are not shown for this estimate, since it is not known how representative Yatir is of the global semi-arid area, nor how representative is the 6 m depth. Forestation related calcite precipitation data in the USZ of other semi-arid regions are not yet available. More accurate estimates can be made once more data are obtained globally on the rate of precipitation of calcium carbonate from the DIC into USZ sediments. A value of 2.5 billion tons $yr^{-1}$ of $CO_2$ removed from the atmosphere represents a respectable ~13% of the present increase of 20 billion tons $yr^{-1}$ of $CO_2$ in the atmosphere.

The carbon global mass balance is not fully constrained (Schlesinger, 2000; Houghton, 2007; Ballantye, et al., 2012; Evans et al., 2014), for there appears to be a terrestrial sink that is lacking. Recently, Li et al. (2015) noted that DIC stored within old saline/alkaline groundwater beneath desert regions may contribute to constraining the carbon budget. This suggestion is consistent with our previous estimates (Moinester et al., 2014; Godfrey-Smith et al., 2015); though we considered two factors in the quantification of the estimated missing terrestrial sink: inorganic carbon storage precipitated in the unsaturated zone (USZ), as well as DIC in global aquifers.

**5. Conclusions**

From carbon isotope data of the DIC, it can be seen that the source of carbon dioxide in the carbonate originates from the atmosphere. Thus, a net sequestration of atmospheric $CO_2$ occurs when carbonate is precipitated. The field data indicates that Eqs. 1-3 must be interpreted with caution. In the field, the dissolution, precipitation of carbonate, and release of $CO_2$ are complex processes. Other sources of calcium should be expected to combine within the high partial pressure of the soil-gas. These sources could include the presence of doubly charged cations adhering to the surface of clay minerals, or the presence of non-carbonate calcium bearing minerals such as gypsum, and sea spray aerosols. The precipitation of inorganic carbon from the DIC of the soil moisture is identified, and its $CO_2$ source traced to the atmosphere. The sequestration is not only recognizable, its rate is quantifiable, knowing the flow rate of water (using tritium in HTO as a tracer) through the USZ. The sequestration can be initiated by all rooting plants that exhale $CO_2$ into the soil zone. Seasonal plants only work for several month of the year, and have shallow roots. Planted forests emit $CO_2$ into USZ throughout the year, and have deeper roots.

Until now, the amount of pedogenic calcite could not easily be discerned from the allogenic limestone, where the two are admixed, which is generally the case. Only through isotopic analyses could ratios between sequestered pedogenic inorganic calcite and primary inorganic limestone be inferred. The reason is that marine limestone has a $\delta^{13}C$ value of approximately 0‰; while calcite formed from the exhalations of $CO_2$ from C3 plants has a $\delta^{13}C$ of approximately -12‰.

The sequestration of atmospheric $CO_2$ by inorganic carbonates in semi-arid regions will not impinge on the food budget. It will not take productive agricultural land and replace it with forests, as might happen in temperate zones. This suggests that forestation projects in semi-arid regions would store atmospheric carbon not only as organic carbon, but also as inorganic carbon. Moreover, the inorganic carbon should be considered as a net long-term sequestration mechanism.

The inorganic storage discussed here is most relevant to semi-arid regions. Far from being a drawback, this gives added economic potential to these marginal regions. Advantages of forestation include employment replacing subsistence agriculture or herding, as well as providing a long-term underground storage of carbon, even after the forests are culled for timber.

While the ideal chemical reactions (Eqs. 1-3) suggest that pedogenic calcite formation within the USZ does not provide a net sequestration gain for atmospheric $CO_2$, the field data demonstrates otherwise. This is our primary conclusion. It leads to the next conclusion regarding the efficacy of forestation as a mode of attenuating atmospheric $CO_2$ emissions. Forestation in semi-arid regions not only increases the biomass organic carbon, but would provide long-term sequestration as insoluble pedogenic carbonate within the USZ. The *"gedanken"* estimate here suggests a significant potential to achieve long-term sequestration of inorganic carbon in semi-arid regions globally.

To date, most semi-arid regions are not industrialized, and are too dependent on rainfall to offer more than employment in only marginally profitable agriculture or herding. Afforestation and reforestation however can provide long-term sustainable employment to such populations at less initial and continuing running costs than required for major engineering endeavors for fulfilling climate goals. Moreover, the planting of trees may also have positive climate cooling effects and prevent encroaching desertification. Besides, over the years, the above ground biomass carbon (organic carbon) can be harvested to supply such commodities as lumber and charcoal. This is not merely speculation. Newly independent Israel suffered similar problems. Forestation efforts were carried out initially to provide employment as a stopgap effort, until factories and suitable jobs could be established. The social fabric of the country was helped by planting forests in a country of limited natural resources, before the global climate aspect was a consideration. The forests today are esthetic, counter encroaching desertification, and provide wild life habitat and recreational facilities. Other semi-arid economically distressed regions, such as the Sahel, where the sediment is also loess, should be similarly forested; but on a large scale with international assistance at first. In this case, the primary aim would be carbon dioxide sequestration in the form of inorganic carbon.

The ongoing major reforestation projects, the Green Walls of China, India, and the Sahel, should reduce atmospheric carbon levels, not only by storage of organic carbon but through the precipitation of pedogenic carbonate in semi-arid regions. By analogy with our study sites, we infer a potential for net carbon removal. Extrapolating our rough estimates for carbon removal rates to global semi-arid environments, there is a potential of removing 5-20% of the current anthropogenic annual $CO_2$ increase to the atmosphere. The data suggest that forestation of these regions may yield significant long-term  inorganic sequestration, in addition to short-term organic carbon sequestration.

Table 1. Data from the Yatir Forest shallow soil core profile.

| Depth (cm) | Humidity (%) | DIC (mmol $CL^{-1}$) | $\delta^{13}C$ (‰) | $\delta^{18}O$ (‰) |
|---|---|---|---|---|
| 30-60 | 10.7 | 4.5 | -15.6 | 0 |
| 60-90 | 12.0 | 3.4 | -14.4 | -2 |

Fig.1a.

  Nizzanim site showing the decrease in DIC down profile.

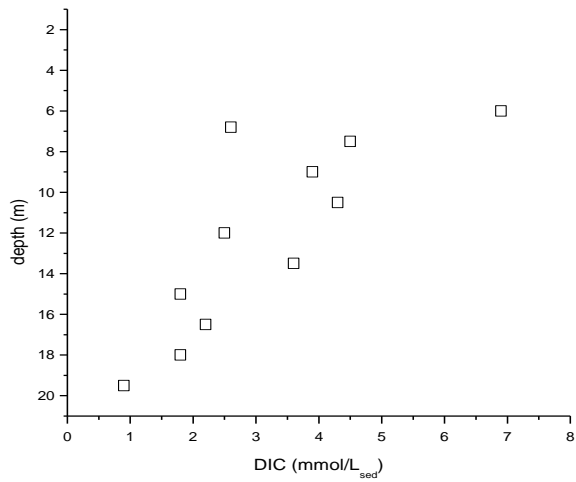

Fig. 1b.

The corresponding decrease of the $^{14}$C in the DIC is not due to radioactive decay. The time is much too short. Rather, there is a continuous carbon isotopic exchange and precipitation that occurs between the DIC and the solid carbonate. (Figs. 1a & b, from Carmi et. al. (2013))

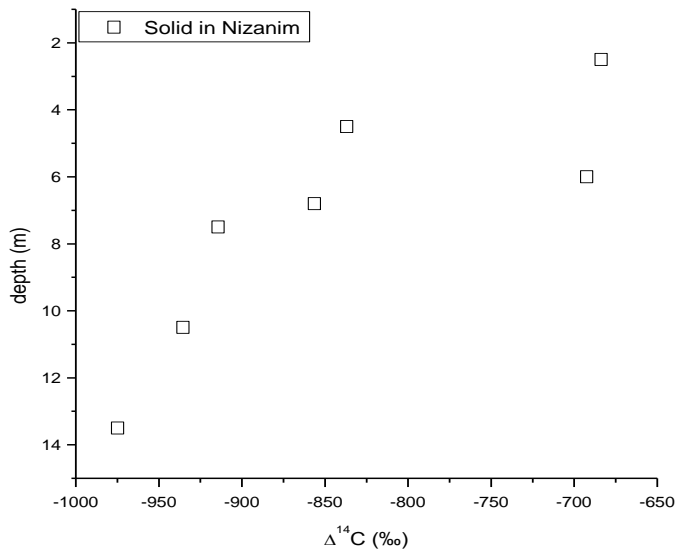

Fig. 2.

The compressed Yatir section illustrates the exchange between the stable carbon isotopes (expressed as $\delta^{13}C$) between the DIC and the relict marine carbonate within the sediment. The allogenic limestone (relict marine carbonate) has a $\delta^{13}C$ of approximately 0‰, which becomes depleted by exchange with the DIC. The latter represents the root exhalations of $CO_2$ from C3-type plants ($\delta^{13}C = \sim -26‰$). Likewise, the relict marine limestone, which has no initial radiocarbon ($\Delta^{14}C = -1000‰$), gains radiocarbon through exchange with the DIC. This results in a concomitant decrease of radiocarbon in the DIC, considering its transfer to the solid. The radiocarbon which is created in the upper atmosphere is transferred to the DIC primarily via the root exhalations of $CO_2$, along with any decomposition of organic material in the soil of the USZ. The $\Delta^{14}C$ traces the incorporation of atmospherically derived $CO_2$ into the carbonate as the pedogenic calcite precipitates from the DIC.

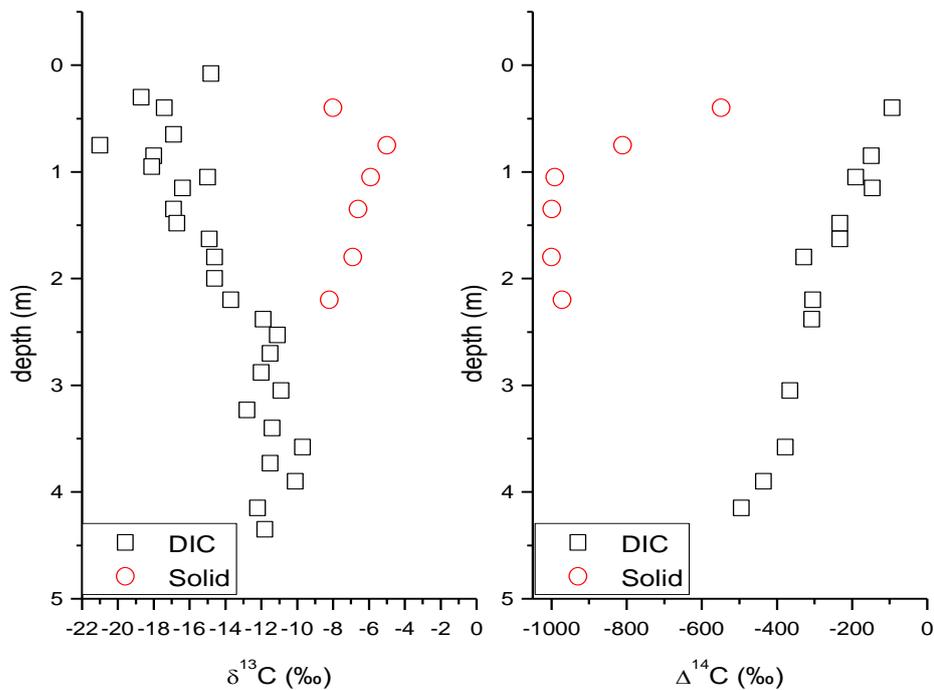